\documentstyle[prl,twocolumn,aps,floats,epsf,epsfig,amsmath,amssymb]{revtex}
\begin{document}

\twocolumn[
\hsize\textwidth\columnwidth\hsize\csname@twocolumnfalse\endcsname
\draft

\title{Transport properties of two-dimensional electron systems on
  silicon (111) surfaces}
\author{E. H. Hwang and S. Das Sarma}
\address{Condensed Matter Theory Center, 
Department of Physics, University of Maryland, College Park,
Maryland  20742-4111 } 
\date{\today}
\maketitle

\begin{abstract}

We theoretically study transport properties of a two-dimensional electron
system on a hydrogen-passivated Si(111) surface in the
field-effect-transistor (FET) configuration. We calculate
the density and temperature dependent mobility
and resistivity for the recently fabricated Si(111)-vacuum FET
by using a semiclassical Boltzmann theory
including screened charged impurity
scattering. We find reasonable agreement
with the corresponding experimental transport properties, indicating
that the screened disorder
potential from random charged impurities is the main
scattering mechanism. We also find that the theoretical results with
the valley degeneracy $g_v=2$ give much better agreement with experiment
than the $g_v=6$ situation indicating that the usual bulk six-valley
degeneracy of Si is lifted in this system.


\end{abstract}
\vspace{0.5cm}
]

\newpage

Recently a high mobility two dimensional electron system (2DES)
on Si(111) surface has been fabricated \cite{Eng,Eng1}. 
The fabricated Si-vacuum
field-effect transistor (FET), in which an electric field is applied
through an encapsulated vacuum cavity instead of the usual SiO$_2$ as
in a Si MOSFET, allows the 2DES to be gated on a 
hydrogen-passivated Si(111) surface.
The strong scattering potential associated with the disordered
Si-SiO$_2$ interface in the usual Si MOSFET geometry \cite{Ando} being
absent at the H-passivated Si(111)-vacuum interface, the
carrier mobility is expected to be higher in this new 2D system
compared with the extensively-studied \cite{Ando} Si(111)-SiO$_2$ 2D
MOSFET system. In the usual Si-SiO$_2$ MOSFETs the carrier mobility is
limited by charged impurity scattering at low carrier densities, but
at higher carrier densities, when the 2DES is pushed very close to the
Si-SiO$_2$ interface by the steep self-consistent confining potential
arising from the carriers themselves, surface roughness and defect
scattering associated with the 
Si-SiO$_2$ interface becomes dominant, reducing
carrier mobility with increasing density. The 2D mobility in
Si-SiO$_2$ MOSFETs therefore has a peak value at some intermediate
carrier density ($\sim 10^{12}$ cm$^{-2}$) with the mobility limited by
charged impurity scattering (interface roughness scattering) below
(above) the characteristic density where the 2D mobility attains its
peak value. Potential advantages of the newly created
\cite{Eng,Eng1} Si-vacuum FET structure are that the absence of
interface roughness/dangling bond scattering leads to (1) very high 2D
mobilities not achievable in Si-SiO$_2$ MOSFET structures; and (2) the
mobility should, in principle, be a monotonically increasing function
of 2D carrier density, as, for example, in 2D undoped GaAs HIGFET
structures \cite{Kane}.
In the Si-vacuum FET fabricated by Eng {\it et al.} \cite{Eng,Eng1}
the H-Si(111) surface reduces surface
scattering, thereby increasing electron
mobility. The measured \cite{Eng}
peak electron mobility on H-Si(111)-vacuum FET
reaches $\sim8000$ cm$^2$/Vs at 4.2K, which is the highest electron
mobility recorded on a Si(111) surface compared to the previous peak electron
mobility measurements in Si(111) MOSFETs ($\sim2500$ cm$^2$/Vs)
\cite{Tsui,Tsui1,Cole}.  With this enhanced electron mobility, the
Si-vacuum FET has potential development of atomic-scale
devices, where electrons are coupled to specific molecules
or single atoms positioned on the H-passivated Si surface.
Indeed, with further improvement in the surface passivation techniques
developed in refs. \onlinecite{Eng,Eng1}, the experimental
mobilities should go up much higher than the currently published
\cite{Eng} record of $\sim 8000$ cm$^2$/Vs.

In contrast to the well-studied electron inversion layers in Si(100)
MOSFETs \cite{Ando}, the detailed investigation of electronic transport
properties of 2DES on Si(111) surface has not been reported in the
literature. Since the high mobility FET on Si(111)
surface is now available, a detailed study of electronic transport
properties is necessary. The purpose of this paper is to
calculate the density and temperature dependent electron
mobility of the 2DES on
Si(111) surface. We also calculate
the 2D temperature-dependent resistivity in connection with the 
subject of 2D metal-insulator transition and the 2D
metallic phase which has been studied for most other high mobility 
2D systems \cite{dsd_ssc}.
The measured temperature dependent resistivity in Si-vacuum FET shows
strong metallic behavior \cite{Eng1} (i.e. the resistivity increases
with temperature) with no sign of insulating
behavior down to a density $n=3.7 \times 10^{11}cm^{-2}$ (which is
the lowest achieved experimental density so far).
The density dependence of mobility  in the
metallic regime of the Si-vacuum FET may give valuable information about the
scattering mechanisms operational in limiting 2D carrier
transport, since different scattering sources 
give rise to different density dependence in the resistivity. 
Thus, a detailed study of the density and temperature 
dependence of 2D resistivity in high-mobility 2D Si(111)-vacuum FET
could be useful not only for a better fundamental understanding of the
limiting transport mechanisms in this potentially important new
system, but also for providing ideas about how to further increase the
2D mobility leading to better devices.

In order to calculate the low temperature 2D electron mobility
we have carried out a microscopic transport calculation using the
Boltzmann transport theory \cite{dsd_ssc}. 
We calculate the mobility in the presence of randomly distributed
Coulomb impurity charges 
near the surface and the bulk-acceptors in the depletion layer
with the electron-impurity  
interaction being screened by the 2D electron gas in the random 
phase approximation (RPA). 
The screened Coulomb scattering  is
the only important scattering mechanism in our calculation.
There are additional interface-scattering mechanisms unrelated to the
Coulomb centers (e.g. surface roughness scattering), but such interface
scattering would become more pronounced at 
higher densities, as the inversion layer is drawn closer to the
interface. In general, the surface roughness scattering gives rise to
the decrease of the mobility at high densities. 
The observed mobility \cite{Eng} shows that the mobility increases
monotonically with electron density indicating the apparent absence of
interface scattering, as expected for a Si-vacuum interface.
In fact, as argued above in this paper, the surface roughness
scattering should be negligible in the Si-vacuum FET systems, allowing
us to ignore this process in our theory.
We also neglect all phonon scattering effects 
mainly because our theoretical
estimate shows phonon scattering to be negligible for Si-vacuum FET 
structures in the $T<10 K$ regime of interest to us. 
We note that our theoretical approach has 
provided a reasonable 
description of the  transport properties for several 2D systems
\cite{dsd_ssc}.

In the Boltzmann theory the
mobility is given by $\mu = e \langle \tau \rangle/m$ with 
$m$ as the carrier effective mass, and $\langle \tau \rangle$  the
energy averaged finite temperature scattering time:
$\langle \tau \rangle = {\int d\epsilon_k \epsilon_k \tau(\epsilon_k)
  \left ( -\frac{\partial 
      f}{\partial \epsilon_k} 
\right )}/{\int d\epsilon_k \epsilon_k \left ( - \frac{\partial
    f}{\partial \epsilon_k} \right )}$, 
where $f(\epsilon_k)$ is the Fermi distribution function,
$f(\epsilon_k) =\{ 1+\exp[  
(\epsilon_k-\mu)]/k_B T \}^{-1}$ with $\mu(T,n)$ as the finite
temperature chemical  
potential determined self-consistently. 
The energy dependent scattering time $\tau(\epsilon_k)$ for our model of 
randomly distributed impurity charge centers is given in the
leading-order theory by
\begin{eqnarray}
\frac{1}{\tau(\epsilon_k)} & = & \frac{2\pi}{\hbar}\sum_a \int n_i^{(a)}(z)
dz \int\frac{d^2k'}{(2\pi)^2}
\left |\frac{v(q,z)}{\varepsilon(q)}\right |^2 \nonumber \\ 
& & \times (1-\cos\theta) \delta\left (
\epsilon_{\bf k} - \epsilon_{\bf k'} \right ),
\end{eqnarray}
where $n_i^{(a)}$ is the concentration of the $a$-th kind of charged
impurity 
center,  $q = |{\bf k} - {\bf k}'|$, $\theta \equiv \theta_{{\bf kk}'}$
is the scattering angle between the scattering in- and out- wave
vectors ${\bf k}$ and ${\bf k}'$, 
$\epsilon_{\bf k} = \hbar^2k^2/2m$, $v(q,z)$ is the 2D Coulomb 
interaction between an electron and an impurity at $z$, and $\varepsilon(q) 
\equiv \varepsilon(q;\mu,T)$ is the 2D finite temperature static RPA 
dielectric (screening) function \cite{Ando}.
In calculating the Coulomb interaction and the RPA dielectric function 
in Eq. (1) we take into account subband quantization effects in the 
inversion layer through the lowest subband variational wavefunction 
\cite{Ando}. From calculated mobility we get the resistivity given by
$\rho =  1/ne\mu$.


Throughout this paper we use the following material parameters: valley
degeneracy $g_v=2$ (or $g_v=6$ as the case may be), an 
effective mass $m=0.35m_e$ (corresponding to the Si(111) surface)
where $m_e$ is the free  
electron mass,  $\kappa_{Si}=11.7$ and $\kappa_{vac}=1$, the dielectric
constant of Si and vacuum, respectively.
Even though the valley degeneracy ($g_v$) on the Si(100) surface is
well-known to be $g_v=2$, the valley degeneracy on the Si(111) surface
has been a puzzle for a long time \cite{Ando}.
Since Si has six conduction-band minima located along the (100) directions,
the (111) surface should have six equivalent valleys (i.e. $g_v=6$)
within the simple effective mass approximation.
However, 
magnetotransport measurements have found the valley degeneracy
to be both 2 and 6 \cite{Tsui,Tsui1,Cole,Ando}, with most experiments
finding the lower valley degeneracy of 2. 
In our calculation we get much better agreement with the experimental
data of refs. \onlinecite{Eng,Eng1}
when we use the valley degeneracy $g_v=2$ instead of $g_v=6$,
indicating that the valley degeneracy is reduced from 6 to 2 in the
Si(111)-vacuum FET structures of ref. \onlinecite{Eng}.

\begin{figure}
\epsfysize=2.3in
\centerline{\epsffile{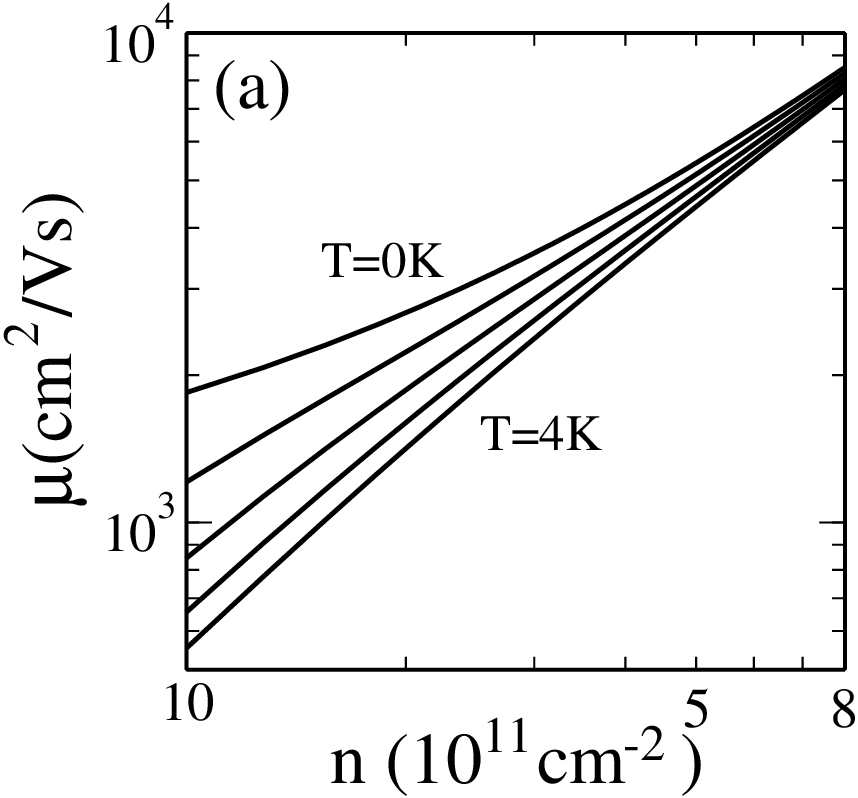}}
\centerline{\epsffile{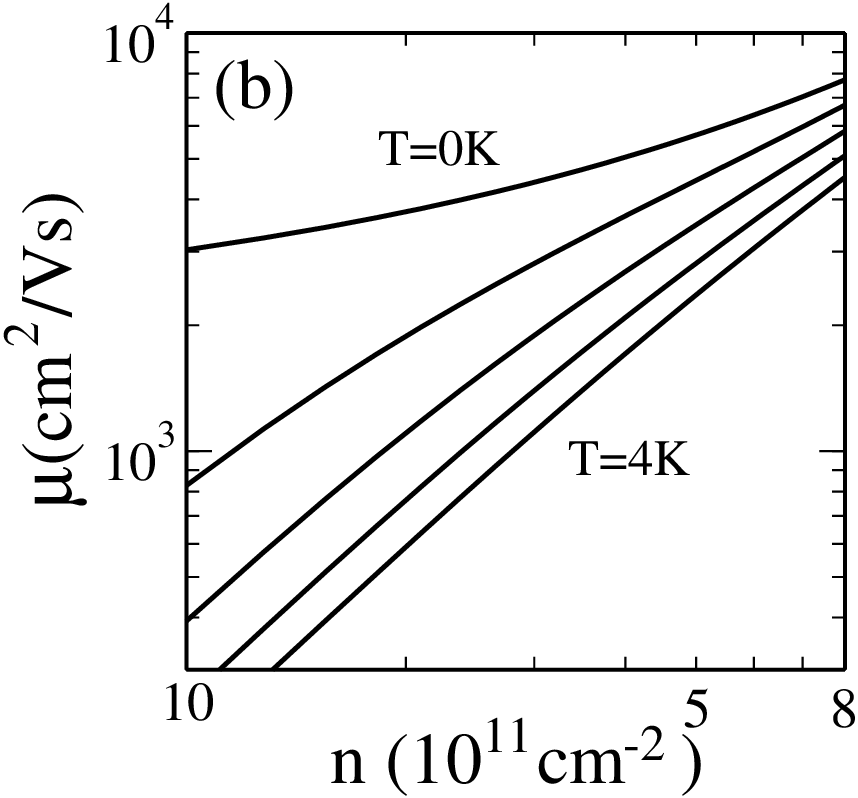}}
\centerline{\epsffile{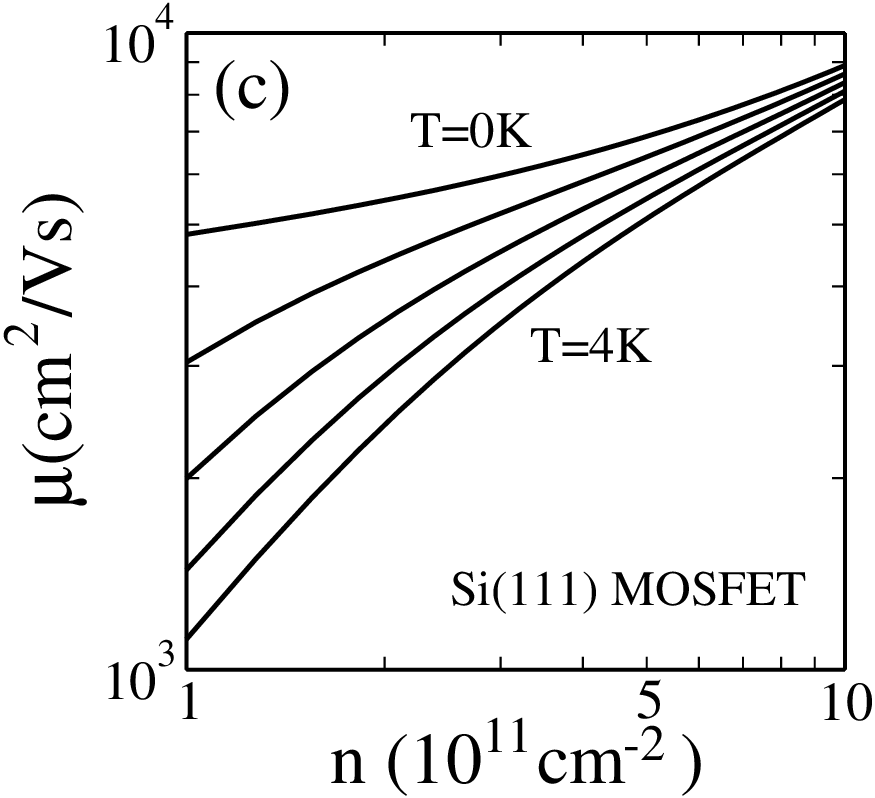}}
\caption{
Calculated electron mobility in the Si(111)-vacuum FET system for
various temperatures, $T=0$, 1, 2, 3, 
4K (top to bottom) with valley degeneracy (a) $g_v=2$ and (b) $g_v=6$.
Fig. 1(c) shows the calculated electron mobility for a Si(111)-SiO$_2$ 
MOSFET with $g_v=2$.
The calculated mobilities depend on the density approximately as $\mu
\sim n^{\alpha}$ with $\alpha \approx 0.7$ at $T=0$ and  $\alpha
\approx 1.2$ at $T=4K$ for $g_v=2$ [Fig. 1(a)], and $\alpha \approx
0.5-1.5$ for $g_v=6$ [Fig. 1(b)]. For Si(111) MOSFET [Fig. 1(c)] we
have $\alpha \approx 0.3-0.6$. Note that surface roughness scattering,
known to be important for Si-SiO$_2$ MOSFETs at higher carrier
densities, is neglected in the calculation.
}
\label{fig_mu}
\end{figure}

In Fig. \ref{fig_mu} we show the calculated mobility as a function of
electron density for different temperatures. 
The mobilities for Si(111)-vacuum FET are shown in
Fig. \ref{fig_mu}(a) with valley degeneracy $g_v=2$ and
Fig. \ref{fig_mu}(b) with $g_v=6$. In Fig. \ref{fig_mu}(c) we show the
mobility of a Si(111) MOSFET system with $g_v=2$. 
For the Si-SiO$_2$ MOSFET calculations we use $\kappa_{SiO_2} = 3.9$
with the other parameters being the same as in the Si-vacuum system.
In Fig. \ref{fig_mu} the mobilities are calculated with fixed
impurity density 
$n_i=7\times10^{11}$ cm$^{-2}$ (the impurity centers are distributed
completely at random in a 2D plane at $d=50$\AA) and the ionized impurity
density $n_{3D} = 4 \times 10^{16}$ cm$^{-3}$ in the depletion layer of
the $p$-type Si.
Here $d$ indicates the distance between the
impurity charge centers and the inversion layer.
This is intended to represent the
average distance between an electron in the inversion layer and the
impurity near the Si-vacuum interface. 
The calculated mobilities depend on the density approximately as $\mu
\sim n^{\alpha}$ for the calculated density range. We emphasize that this
is not a strict power law 
since the exponent $\alpha$ depends weakly on the density. 
Over the experimental density range,
we find that for $g_v=2$ [Fig. 1(a)]
the exponent $\alpha$ in $\mu \propto n^{\alpha}$ is about 
0.7 at T=0K and increases as temperature increases. At $T=4K$ we have
$\alpha \approx 1.2$ which is consistent with experiment
\cite{Eng,Eng1}.
For $g_v=6$ [Fig. 1(b)] we have $\alpha \approx 0.5-1.5$. Note that the
Si(111) MOSFET system [Fig. 1(c)] has $\alpha \approx
0.3-0.6$ in the 
given temperature range. [The characteristic separation ``d'' between
the charged impurities and the 2DES as well as the random impurity
densities $n_i$ and $n_{3D}$ are of course not precisely known for the
experimental samples, but the parameters we use are quite typical, and
give decent agreement with the experimental results of
ref. \onlinecite{Eng} for $g_v=2$, i.e. Fig. \ref{fig_mu}(a).]
Our mobility calculation indicates
that the screened potential from random charged impurities is the main
scattering source and the Si(111)-vacuum FET has a valley degeneracy
$g_v=2$. Since there
are no adjustable parameters in our calculation except the impurity
parameters, the agreement is quite good. 
The calculated mobility obviously depends on the distance ($d$) of
the charged impurity centers from the 2D layer. Generally the location
of the charged 
impurities in the FET is not known. Thus, the choice of  $d=50$ \AA
\;\;  is reasonable because our calculated mobility has
the same dependence on $n$ as does the measured mobility \cite{Eng,Eng1}.

\begin{figure}
\epsfysize=2.3in
\centerline{\epsffile{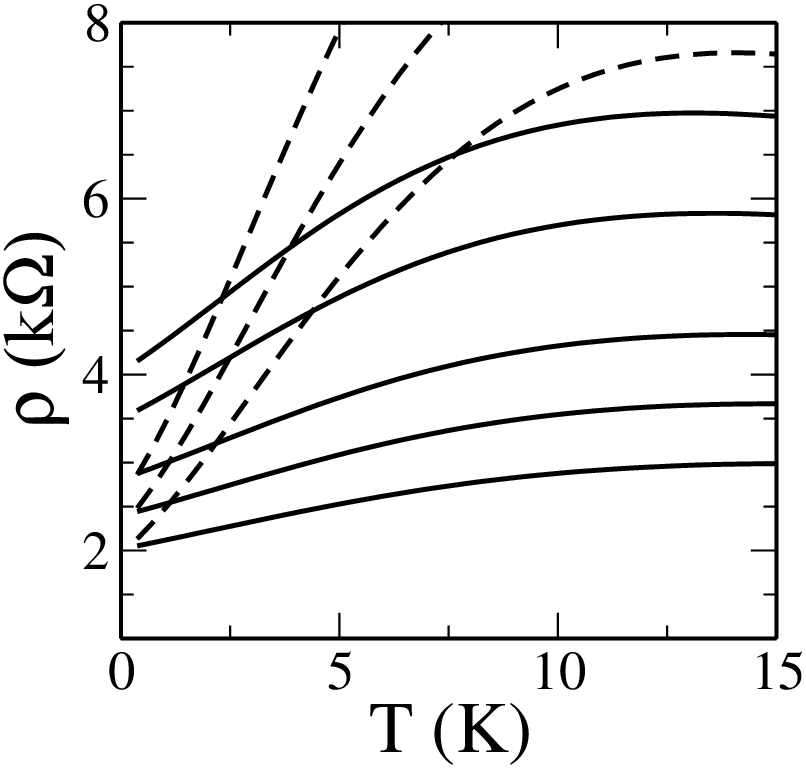}}
\caption{ Calculated resistivities for various hole
densities $n=$3.7, 4.0, 4.5, 4.9, 5.36$\times
10^{11}cm^{-2}$ (from top to bottom) with the same parameters of
Fig. \ref{fig_mu}. The dashed lines show the calculated resistivity for
density $n=$4.5, 4.9, 5.36$\times 10^{11}cm^{-2}$ with the valley degeneracy
$g_v=6$. }
\label{fig_rho}
\end{figure}


In Fig. \ref{fig_rho} we show the calculated temperature dependent
resistivity using the 
same parameters as in Fig. \ref{fig_mu}.
The calculated resistivity increases as the temperature increases
(i.e. metallic behavior), and
shows an {\it approximately} linear
temperature dependence in the $T < 5$K regime as observed in the
experiment \cite{Eng1}. The temperature
dependence of $\rho(T)$ becomes stronger with decreasing
density just as the experimental observation in Si(111)-vacuum FET
\cite{Eng1}. This strong linear temperature
dependence of $\rho(T)$ in this system arises \cite{dsd_ssc,stern}
due to the peculiar nature of the 2D screening function which
has a cusp at wave vector $q=2k_F$. Since $2k_F$ scattering is
the most dominant resistive scattering at low temperatures, thermal
smearing of this $2k_F$ cusp leads to
strong temperature dependence of resistivity. 
Thus, temperature
dependent screening of charged impurity scattering in 2D layers
leads to a strongly temperature dependent
effective disorder controlling $\rho(T,n)$ at low
temperatures and densities. (The screening effect 
decreases with increasing temperature giving rise to
increasing effective disorder with increasing T, and hence
increasing $\rho(T )$ with T.) 
Note that the overall resistivity scale depends on
the unknown impurity densities, but the qualitative trends in
$\rho(T,n)$ arise from basic aspects of the underlying scattering
mechanisms.

We pointed out earlier \cite{dsd_ssc}
that the requirements for the  observation of a large
temperature-induced change in resistivity is 
the strong screening condition, $q_{TF} \gg 2k_F$, where $q_{TF}$ and
$k_F$ are the Thomas-Fermi screening wave vector and Fermi wave
vector, respectively.  Since $q_{TF}/k_F \propto g_v^{3/2}$ we 
expect the observed metallicity to be much stronger for $g_v=6$
than for $g_v=2$.
The dashed lines in the Fig. \ref{fig_rho} show the calculated
resistivity for $n=$4.5, 4.9, 5.36$\times 10^{11}cm^{-2}$ with $g_v=6$.
We find, in the range of of $0-5$K, roughly a factor of
three change in $\rho(T)$ with $g_v=6$  whereas
more like a 50\%  temperature induced
change in $\rho(T)$ with $g_v=2$, which is consistent with the
experimental observation. Thus, the transport measurement in
Si(111)-vacuum FET supports the Si(111) surface having a
valley degeneracy 2.  

Finally, we briefly discuss the possible origin of the valley
degeneracy puzzle in this system, i.e. why $g_v=2$ in the
Si(111)-vacuum FET system rather than the usual effective mass
prediction of $g_v=6$. This is, in fact, an old problem \cite{Ando}
much discussed in the context of Si(111)-SiO$_2$ MOSFETs. It is
generally accepted \cite{Tsui,Tsui1} that the random strain associated
with the Si 
surface lifts the valley degeneracy lowing two of the six valleys in
energy. Only specially processed Si(111)-SiO$_2$ samples with rather
low mobilities were ever found to have $g_v=6$ \cite{Tsui1} with most
Si(111)-SiO$_2$ MOSFETs having $g_v=2$. We believe the same kind of
one electron physics to be operational in lifting the valley
degeneracy from $g_v=6$ to $g_v=2$ in the Si(111)-vacuum system as
well  although understanding the details of this valley degeneracy
puzzle is beyond the scope of our current work where we are interested
in the transport properties. We do, however, mention that this valley
degeneracy lifting is unlikely to be a many-body exchange instability,
which would occur even as a matter of principle at a much lower critical
density ($\sim 10^{11}$ cm$^{-2}$) with a critical temperature of only 1K
or so. At the experimental density and temperature range of
ref. \onlinecite{Eng}, a many-body effect induced valley degeneracy
lifting is extremely unlikely \cite{Beni,Sankar}

In conclusion, we calculate the
the density and temperature dependent mobility
and resistivity for the recently fabricated Si(111)-vacuum field
effect transistor by using a semiclassical Boltzmann theory
including screened charged impurity
scattering. We find reasonable agreement
with the corresponding experimental transport properties for the (111)
system indicating
that the screened potential from random charged impurities is the main
scattering source. 
In comparing our calculation with the existing experimental data
\cite{Eng,Eng1} we find that the calculated results with
the valley degeneracy $g_v=2$ give better agreement with experiments
than $g_v=6$.

This work is supported by LPS.

\end{document}